\documentclass[aps,preprint,amsmath,amssymb,pra]{revtex4-2}
\usepackage{graphicx}% Include figure files
\usepackage{bm}% bold math
\begin{document}
%\preprint{APS/123-QED}
\title{Farey tree locking of terahertz semiconductor laser frequency combs}
\author{Guibin Liu$^{1,2}$, Xuhong Ma$^{1,2}$, Kang Zhou$^{1,4}$, Binbin Liu$^{1,2}$, Lulu Zheng$^{1,2}$, Xianglong Bi$^{1,2}$, Shumin Wu$^{1,2}$, Yanming Lu$^{1,2}$, Ziping Li$^{1}$, Wenjian Wan$^{1}$, Zhenzhen Zhang$^{1}$, Junsong Peng$^{3}$,Ya Zhang$^{5}$}
\author{Heping Zeng$^{3,4,*}$}
%\email{}
\author{Hua Li$^{1,2,}$}
\email{Corresponding author: hpzeng@phy.ecnu.edu.cn\\
Corresponding author: hua.li@mail.sim.ac.cn.}

% \author{Guibin Liu,$^{1,2}$ Xuhong Ma,$^{1,2}$ Kang Zhou,$^{1,4}$ Binbin Liu,$^{1,2}$, Lulu Zheng,$^{1,2}$ Xianglong Bi,$^{1,2}$ Shumin Wu,$^{1,2}$ Yanming Lu,$^{1,2}$ Ziping Li,$^{1}$ Wenjian Wan,$^{1}$ Zhenzhen Zhang,$^{1}$ Junsong Peng,$^{3}$ Ya Zhang,$^{5}$}
% \author{Heping Zeng,$^{3,4}$}
% \email{Corresponding author: hpzeng@phy.ecnu.edu.cn}
% \author{Hua Li,$^{1,2}$}
% \email{Corresponding author: hua.li@mail.sim.ac.cn.}

\affiliation{%
$^{1}$State Key Laboratory of Materials for Integrated Circuits and Key Laboratory of Terahertz Solid State Technology, Shanghai Institute of Microsystem and Information Technology, Chinese Academy of Sciences, 865 Changning Road, Shanghai 200050, China.\\
$^{2}$Center of Materials Science and Optoelectronics Engineering, University of Chinese Academy of Sciences, Beijing 100049, China.\\
$^{3}$State Key Laboratory of Precision Spectroscopy, East China Normal University, Shanghai 200241, China.\\
$^{4}$Chongqing Key Laboratory of Precision Optics, Chongqing Institute of East China Normal University, Chongqing 401120, China.\\
$^{5}$Institute of Engineering, Tokyo University of Agriculture and Technology, Koganei-shi 184-8588 Tokyo, Japan
}%

%\collaboration{MUSO Collaboration}%\noaffiliation
%\collaboration{CLEO Collaboration}%\noaffiliation

%\date{\today}% It is always \today, today,
             %  but any date may be explicitly specified

\begin{abstract}
Frequency combs show various applications in molecular fingerprinting, imaging, communications, and so on. In the terahertz frequency range, semiconductor-based quantum cascade lasers (QCLs) are ideal platforms for realizing the frequency comb operation. Although self-started frequency comb operation can be obtained in free-running terahertz QCLs due to the four-wave mixing locking effects, resonant/off-resonant microwave injection, phase locking, and femtosecond laser based locking techniques have been widely used to broaden and stabilize terahertz QCL combs. These active locking methods indeed show significant effects on the frequency stabilization of terahertz QCL combs, but they simultaneously have drawbacks, such as introducing large phase noise and requiring complex optical coupling and/or electrical circuits. Here, we demonstrate Farey tree locking of terahertz QCL frequency combs under microwave injection. The frequency competition between the Farey fraction frequency and the cavity round-trip frequency results in the frequency locking of terahertz QCL combs, and the Farey fraction frequencies can be accurately anticipated based on the downward trend of the Farey tree hierarchy. Furthermore, dual-comb experimental results show that the phase noise of the dual-comb spectral lines is significantly reduced by employing the Farey tree locking method. These results pave the way to deploying compact and low phase noise terahertz frequency comb sources. 
% \begin{description}
% \item[Usage]
% Secondary publications and information retrieval purposes.
% \item[Structure]
% You may use the \texttt{description} environment to structure your abstract;
% use the optional argument of the \verb+\item+ command to give the category of each item. 
% \end{description}
\end{abstract}

%\keywords{Suggested keywords}%Use showkeys class option if keyword
                              %display desired
\maketitle

%\tableofcontents

\section{INTRODUCTION}
Frequency combs are generated by mode-locked pulse trains in the time domain and manifest as a series of equidistant spectral lines in the frequency domain\cite{udem2002optical}. The frequency comb offers advantages such as a broad spectrum and high coherence, making it a powerful light source in spectral detection, imaging, and communications\cite{diddams2007,ideguchi2013,marin2017}. In the terahertz (THz) frequency range (0.1-10 THz), semiconductor-based and electrically pumped quantum cascade lasers (QCLs) have been proven to be compact and high power terahertz radiation sources\cite{faist1994,kohler2002, Zhuo2020}. The self-started frequency comb operation in free-running THz QCLs can be obtained due to the four-wave mixing locking mechanism, and its performance can be improved by optimizing the group velocity dispersion (GVD) and/or implementing dispersion compensators in laser structures\cite{burghoff2014,zhou2019,wangfeihu2017}. 

To further improve the stability of THz QCL frequency combs, active locking techniques have been widely employed. For example, the resonant microwave injection which was traditionally used for active mode locking of semiconductor diode lasers\cite{bowers1989} was also successfully adapted for active locking of the repetition frequency ($f_{\rm{rep}}$) of THz QCL combs. By employing the resonant microwave injection technique, the repetition frequency of the QCL can be firmly locked to the injection microwave frequency and the optical spectrum can be significantly broadened\cite{gellie2010,wan2017,hillbrand2019,li2019, Forrer2020}. Moreover, the off-resonant microwave modulation with a frequency detuning of $\sim$200 MHz was also demonstrated to broaden the spectrum of QCL combs by introducing phase matching and then enhancing the nonlinear four-wave mixing locking effect\cite{liao2022}. To further lock the offset frequency of THz QCL combs, phase locking and femtosecond laser based locking techniques have been reported\cite{consolino2019,guan2023,oustinov2010,barbieri2011}. Although the above-mentioned active locking methods show effective stability improvements for THz QCL combs, they simultaneously show obvious drawbacks, such as introducing large phase noise, which is detrimental to the further deployment of dual-comb sources, and requiring complex optical coupling and/or electrical circuits. Therefore, it is urgent to develop a locking technique that is easy to implement and at the same time introduces low phase noise in THz QCL combs. 

The competition between two frequencies in a nonlinear system results in their eventual locking at a particular rational number\cite{winful1986, Freeman2017}, which provides a new approach, i.e., Farey tree locking, to lock the comb's repetition frequency by modulating the laser at a lower frequency. Under the locking conditions, the frequency ratios (or the winding numbers) are always fixed at the rational numbers. The collection of all rational numbers is represented by the Farey tree in math theory. The nodes in the Farey tree, known as Farey fractions, are infinitely generated downward along hierarchies and are always accompanied by the devil's staircase fractal structure\cite{bak1986devil}. The relationship between frequency competition and locking was first observed by Huyghens in the 17th century when a pair of clocks were mounted back-to-back on a wall. Recently, this phenomenon has been observed in additional nonlinear optical systems, such as radio frequency (RF) injection in external-cavity semiconductor lasers\cite{baums1989farey}, picosecond pulse sequence-driven microresonator soliton frequency combs\cite{Xu2020}, and breather lasers\cite{wu2022}. Semiconductor-based QCLs, which possess strong nonlinearity and current modulation properties\cite{riepl2021}, show promising potentials to achieve Farey tree locking by employing the current modulation at a Farey fraction frequency.

Here, we demonstrate the Farey tree locking of a THz QCL frequency comb under microwave injection. By tuning the modulation frequency step by step, the locking of the repetition frequency of the THz QCL comb at Farey fraction frequencies is observed. The winding numbers corresponding to the locking states are a series of Farey fractions within the range between 0 and 1, which can be accurately anticipated based on the downward hierarchy of the Farey tree. The locking bandwidths under the Farey tree locking conditions reveal the commonly observed Arnold tongue\cite{ecke1989,sacher1992} structure in nonlinear optics. By distributing the observed Farey tree locking states along the modulation frequency, 11 plateaus are obtained. The width of each plateau decreases as the corresponding Farey tree hierarchy deepens. It is worth noting that as the measurement precision is improved, a new filial plateau can be discovered between two parental plateaus, validating the fractal structure resembling the devil's staircase. Furthermore, by employing the Farey tree locking method, a THz dual-comb experiment is performed and the result shows that the phase noise level of the dual-comb lines under the Farey tree locking condition is significantly reduced, compared to that measured in free-running mode.

\section{RESULTS}

\subsection{Experimental setup and laser performance}

\begin{figure*}[tb]
\centering
\includegraphics[width=0.95\linewidth]{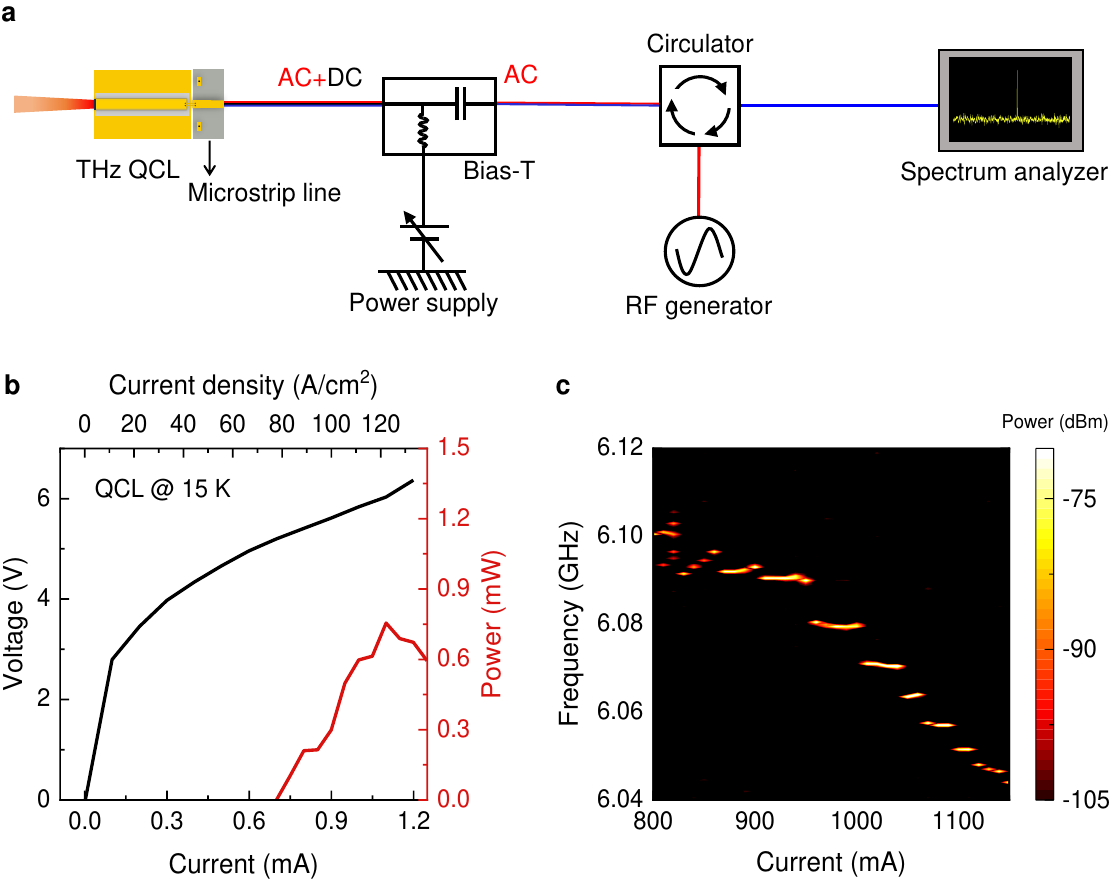}
\caption{Experimental setup and laser performances. (\textbf{a}) Experimental setup. A microstrip line is mounted close to the QCL to facilitate the microwave injection (modulation) and intermode beatnote signal extraction. The microwave circulator is used to isolate the transmissions of microwave injection (red line) and intermode beatnote (blue line) signals. The entire RF signal is finally displayed on the spectrum analyzer. (\textbf{b}) Light–current-voltage curves of the THz QCL measured at a heat sink temperature of 15 K in CW mode. The THz QCL ridge is 150 $\mu$m wide and 6 mm long. (\textbf{c}) Intermode beatnote map of the free-running THz QCL at a heat sink temperature of 15 K, measured with a resolution bandwidth (RBW) of 10 kHz and a video bandwidth (VBW) of 1 kHz.}
\label{setup}
\end{figure*}

Figure \ref{setup}a shows the experimental setup employed for the Farey tree study in THz QCLs under microwave injection. To detect the intermode beatnote signal (or repetition frequency) and confirm its locking condition, the QCL itself is used as a fast THz detector\cite{gellie2010,li2015}. To facilitate the high frequency signal transmission, a microstrip line is mounted close to the QCL chip. The self-detected terminal current signal is transmitted through a microstrip line and the AC port of a Bias-Tee, and then passes through a microwave circulator, which is used to isolate the RF injection signal and intermode beatnote signal. Finally, the electrical intermode beatnote spectrum of the THz QCL comb is displayed by a spectrum analyzer. For the microwave injection, an RF generator is employed to provide stable microwave signals with the expected power and frequency. The microwave signal goes via the circulator, bias-Tee, and microstrip line and is finally injected into the QCL chip.

THz QCL chips are fabricated by using traditional semiconductor laser processing method. The typical light–current-voltage ($L-I-V$) characteristics measured in continuous-wave (CW) mode are shown in Fig. \ref{setup}b. It can be seen that at 15 K the laser can output a maximum CW power of approximately 0.75 mW with a threshold current of 700 mA. Note that the power values shown in Fig. \ref{setup}b are the measured ones displayed on the THz power meter without considering any corrections of water absorption, window transmission, mirror reflections, etc. Figure \ref{setup}c shows the intermode beatnote map of the THz QCL by changing the drive current. One can see that at most current pump conditions ($>$850 mA), single and narrow fundamental intermode beatnotes around 6 GHz are obtained, clearly indicating the frequency comb state. In this work, in order to observe the Farey tree locking states, the QCL is operated in the frequency comb state and a drive current of 1025 mA is chosen.

\subsection{Farey tree in THz QCLs}

\begin{figure}[tb]
\centering
\includegraphics[width=0.98\linewidth]{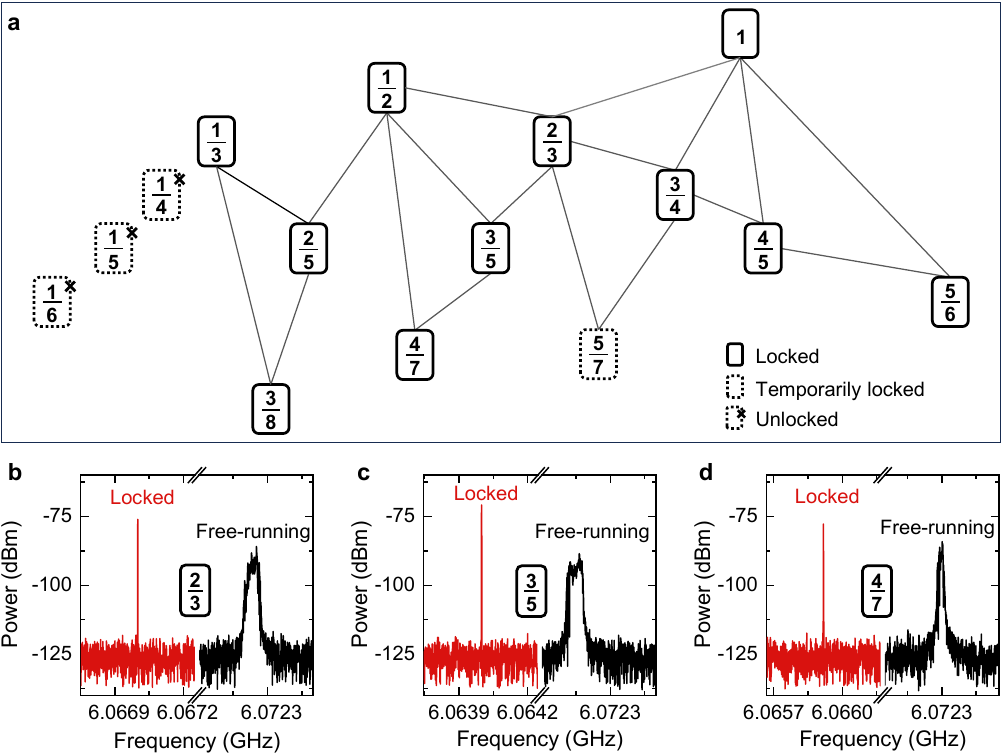}
\caption{Farey tree of the THz QCL frequency comb and typical intermode beatnote spectra under the Farey tree locking conditions. (\textbf{a}) Farey tree structure of the THz QCL comb under microwave injection. The Farey fractions outlined by solid rectangles, dashed rectangles, and dashed rectangles with crosses represent firmly locked, temporarily locked, and unlocked states, respectively. (\textbf{b}), (\textbf{c}), and (\textbf{d}) are measured intermode beatnote spectra at Farey fractions of 2/3, 3/5, nad 4/7, respectively. The red curves are recorded under Farey tree locking conditions, while the black ones are obtained in free-running mode. The RBW and VBW are set as 300 Hz and 30 Hz, respectively. The microwave injection power is 30 dBm.}
\label{farey tree}
\end{figure}

To investigate the Farey tree locking effect in the THz QCL comb, as shown in Fig. \ref{setup}a, we tune the modulation frequency (or injection frequency) generated from the RF generator and monitor the fundamental intermode beatnote signal using a spectrum analyzer. Under the Farey tree locking condition, the Farey fraction is defined as m/n with m being the modulation frequency ($f_{\rm{mod}}$) and n being the fundamental intermode beatnote frequency (or repetition frequency, $f_{\rm{rep}}$) of the THz QCL comb. In this experiment, the ratio m/n is always equal to or smaller than 1. Based on the downward prediction of the Farey-sum operation, new Farey fractions can be generated (see Eq. \ref{eq:median} in Appendix). Figure \ref{farey tree}a shows a part of the Farey tree observed in the THz QCL under microwave injection. We start the microwave injection from the fractions of 1/n located at the top hierarchy of the Farey tree. Initially, the frequency-locked phenomenon is observed under the resonant microwave injection with a Farey fraction of 1/1 ($f_{\rm{mod}}$=$f_{\rm{rep}}$). Then, the locking of $f_{\rm{rep}}$ is also obtained at the Farey fraction of 1/2. A new Farey fraction 2/3 is generated by following the Farey-sum between the known upper-hierarchy fractions of 1/1 and 1/2, and it extends further to deep layers. Following this branch, the Farey tree locking stops at fractions with a denominator of 7. Note that the RF generator used in the experiment cannot output $>$30 dBm power, and therefore, we are not able to go deeper following the Farey tree branch. This can be proved by the temporary locking at Farey fraction of 5/7.

To clearly demonstrate the Farey tree locking effect, in Figs. \ref{farey tree}b, \ref{farey tree}c, and \ref{farey tree}d, we show the locked intermode beatnote spectra at three typical Farey fractions of 2/3, 3/5, and 4/7, respectively, together with the free-running spectra for reference. By comparing the beatnote spectra in the Farey tree locking state with those recorded in free-running, a significant narrowing of the spectral linewidth and an approximate 10 dB improvement in the signal-to-noise ratio (SNR) are observed. This indicates that when modulating the THz QCL at the Farey fraction frequencies, the repetition frequency of the laser comb can be firmly locked. It is interesting to note that when the Farey tree locking takes effect, the first locked $f_{\rm{rep}}$ is always located on the left side of the free-running $f_{\rm{rep}}$. This frequency red-shift phenomenon results from the frequency pulling effect induced by $f_{\rm{mod}}$. As the modulation frequency is increased, the locked $f_{\rm{rep}}$ will then increase in the locking range. 

Returning to the top hierarchy of 1/n and continuing the attempts, a new Farey tree branch with Farey fractions of 1/3 and 1/2 can be obtained, as shown in Fig. \ref{farey tree}a. In this new branch, the deepest layer with a Farey fraction of 3/8 is obtained. In Supplementary Fig. S1a (Supporting Information), we show the comparison of the intermode beatnote spectra recorded in free-running mode and at a Farey fraction of 3/8. Similar to what we observed in Figs. \ref{farey tree}b-\ref{farey tree}d, the linewidth narrowing and SNR enhancement are clearly obtained due to the Farey tree locking. The dynamics of the Farey tree locking process are shown in Fig. S1b (Supporting Information) and a locking range of 90 kHz at the Farey fraction of 3/8 is obtained. It is worth noting that in Fig. \ref{farey tree}a, we also show some fractions outlined by dashed rectangles with crosses. For these winding numbers, the Farey tree locking states cannot be obtained. The absence of frequency locking for these winding numbers reveals that not all rational numbers can induce enhanced nonlinear coupling in frequency competition. It also rules out the possibility that Farey tree locking states arise simply from resonance coupling between higher harmonics of the RF modulation signal and the repetition frequency. It has to be mentioned that in the current experiment, the maximum RF power that we can apply is 30 dBm. If we can further increase the RF power, probably the locking states can be obtained at these Farey fractions.

\begin{figure}[tb]
\centering
\includegraphics[width=0.98\linewidth]{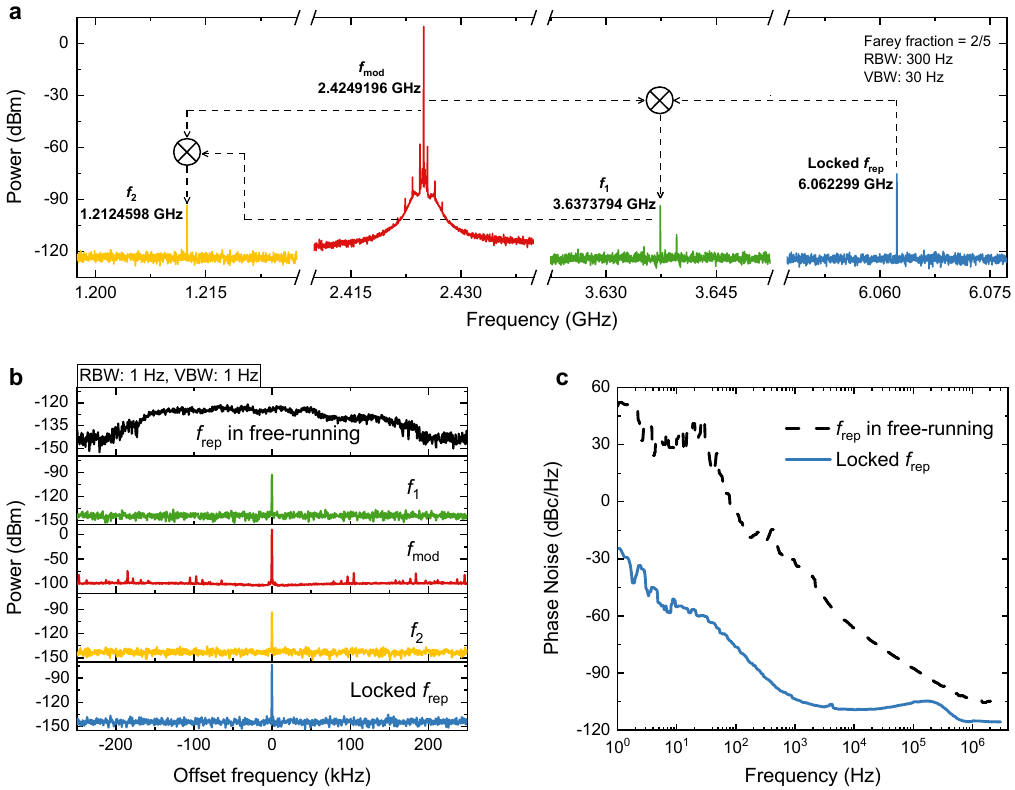}
\caption{(\textbf{a}) Frequency synthesis in microwave frequency range under the Farey tree locking condition with a winding number of 2/5. The RF power of $f_{\rm{mod}}$ is 26 dBm. (\textbf{b}) Spectra of the signals shown in \textbf{a} measured with an RBW of 1 Hz and a VBW of 1 Hz. The spectrum shown in the top panel is the intermode beatnote signal of the THz QCL frequency comb operated in free-running mode for reference. Note that each spectrum is offset by the central frequency for a clear comparison of the linewidth. (\textbf{c}) Phase noise spectra of free-runing $f_{\rm{rep}}$ (dashed line) and locked $f_{\rm{rep}}$ (solid line).}
\label{PN}
\end{figure}

In Fig. \ref{PN}, we present a typical Farey tree locking at a Farey fraction of 2/5. In this situation, the THz QCL is modulated at $f_{\rm{mod}}$=2.4249198 GHz which corresponds to a Farey fraction of 2/5 ($f_{\rm{mod}}$/$f_{\rm{rep}}$). When the Farey tree locking is achieved, $f_{\rm{rep}}$ is firmly locked at 6.062299 GHz, as shown in Fig. \ref{PN}a. Due to the strong nonlinearity in the THz QCL cavity, other frequency components can be observed. For example, $f_{\rm{mod}}$ can beat with $f_{\rm{rep}}$ and generate $f_1$; moreover, $f_1$ mixes with $f_{\rm{mod}}$ and finally generates $f_2$. The mixing processes are schematically shown in Fig. \ref{PN}a. Since $f_{\rm{mod}}$ and $f_{\rm{rep}}$ are stable, the generated $f_1$ and $f_2$ are also very stable. Figure \ref{PN}b plots the beatnote spectra of the four signals, i.e., $f_{\rm{mod}}$, $f_{\rm{rep}}$, $f_1$, and $f_2$, under the Farey tree locking condition, measured with an RBW of 1 Hz and a VBW of 1 Hz. For a clear comparison, the free-running $f_{\rm{rep}}$ is shown in the top panel of Fig. \ref{PN}b. In free-running mode, the broad peak of $f_{\rm{rep}}$ spans over 400 kHz, while once the Farey tree locking is activated $f_{\rm{rep}}$ is immediately locked and a narrow peak with a linewidth of approximately 100 Hz is obtained. Simultaneously, from Fig. \ref{PN}b we can see that $f_1$ and $f_2$ show similar narrow peaks as the locked $f_{\rm{rep}}$, which further proves that the locking is achieved. In Fig. \ref{PN}c, we compare the phase noise spectra of free-running $f_{\rm{rep}}$ (dashed line) and locked $f_{\rm{rep}}$ (solid line). It can be seen that the phase noise of the locked $f_{\rm{rep}}$ is significantly improved. At the offset frequencies of 100 Hz and 10 kHz, the measured phase noise values of locked $f_{\rm{rep}}$ are dramatically reduced by approximately 72 dBc/Hz and 42 dBc/Hz, respectively, compared to those measured in free-running mode. Note that the phase noise spectra are measured using the phase noise analyzer with the low pass filter of a cutoff frequency of 10 MHz (see Appendix).

\begin{figure}[tb]
\centering
\includegraphics[width=0.98\linewidth]{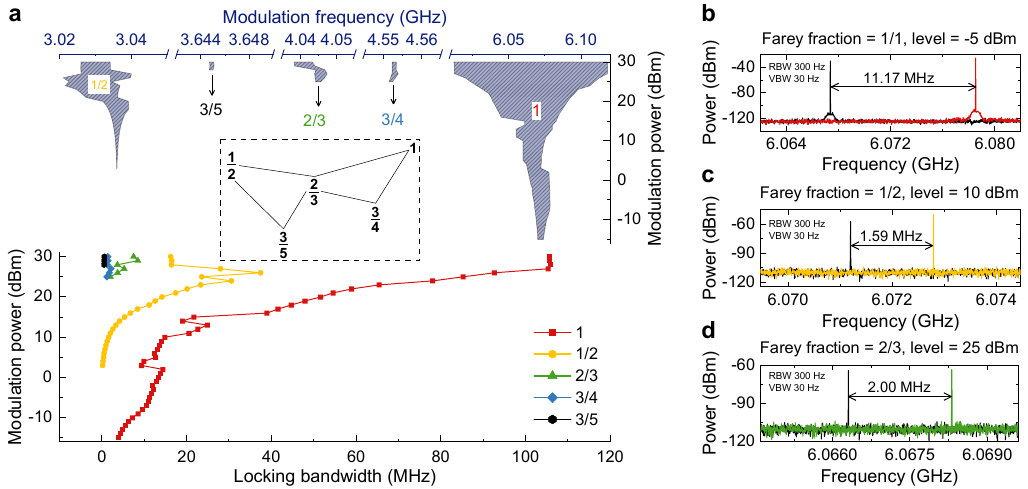}
\caption{Farey tree locking bandwidth. (\textbf{a}) Lower panel (bottom-x and left-y): Modulation power versus locking bandwidth for different Farey fractions outlined by the Farey tree structure in the inset. Upper panel (top-x and right-y): Plots of Arnold tongues corresponding to the results shown in the lower panel. The blue-shaded area represents the Farey tree locking region. In the measurement, the step size for the modulation power is 1 dBm. (\textbf{b}), (\textbf{c}), and (\textbf{d}) are recorded locked intermode beatnote spectra (left and right edges in the locking bandwidth) at Farey fractions of 1/1 (-5 dBm), 1/2 (10 dBm), and 2/3 (25 dBm), respectively.}
\label{Arnold}
\end{figure}

We further investigate the locking bandwidth at typical Farey fractions by changing the modulation power and the main results are shown in the lower panel of Fig. \ref{Arnold}a. One can see that as the denominator of the Farey fraction increases, the threshold modulation power required for the Farey tree locking increases, and at the same time, the achievable locking bandwidth decreases. At 10 dBm RF power, we obtain the locking bandwidths of 14.9 MHz and 1.6 MHz for Farey fractions of 1 and 1/2, respectively. To achieve Farey tree locking for Farey fractions with larger denominators, a large RF power close to the maximum of 30 dBm is needed. The measured maximum locking bandwidths for Farey fractions of 1, 1/2, 2/3, 3/4, and 3/5 are 105.6, 16.2, 7.5, 1.4, and 0.6 MHz, respectively, limited by the achievable RF power of 30 dBm. The characteristics of modulation power versus locking bandwidth can also be depicted by plots of Arnold tongues as shown in the upper panel of Fig. \ref{Arnold}a. The shaded areas indicate the Farey tree locking regions, revealing the Arnold tongues similar to those described in ref. [\citenum{sacher1992}]. In most cases, the tongues demonstrate that as the modulation power increases, the locking bandwidth or its corresponding modulation frequency range also increases. However, it is worth noting that for the Farey fractions of 1/2, a sudden decrease in locking bandwidth is clearly observed as the modulation power exceeds 26 dBm. This could be associated with the nonlinear effects activated in the laser cavity due to the high RF power injection. In Figs. \ref{Arnold}b-\ref{Arnold}d, locked intermode beatnote spectra for three Farey fractions, i.e., 1, 1/2, and 2/3, at different modulation power levels are shown. For each Farey fraction, two typical spectra at the left and right locking boundaries are displayed and the distance between the two peaks indicates the locking bandwidth. 

\begin{figure}[tb]
\centering
\includegraphics[width=0.98\linewidth]{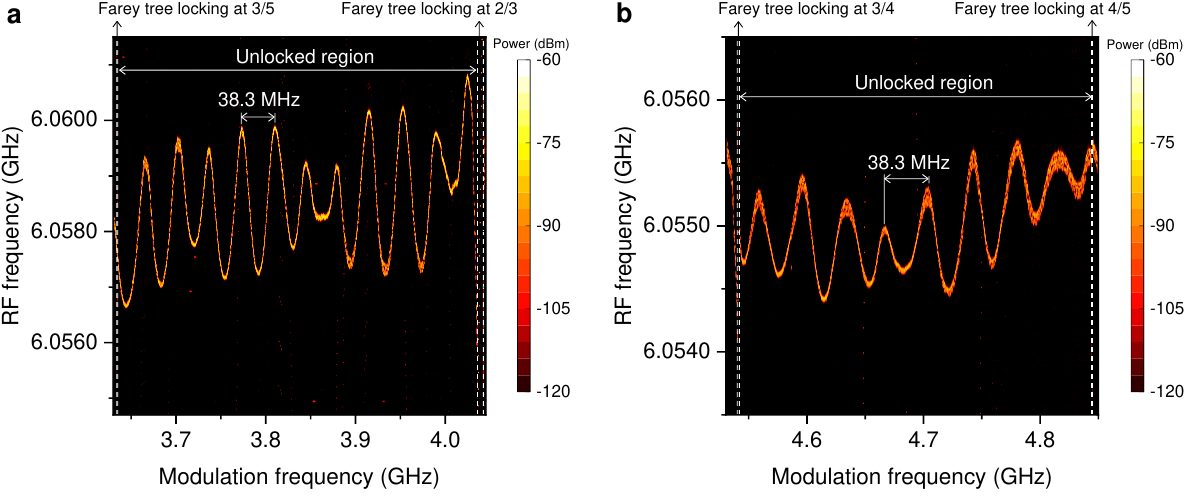}
\caption{Oscillation of repetition frequency with modulation frequency. (\textbf{a}) Repetition frequency map measured in the unlocked region between two Farey tree locking states with Farey fractions of 3/5 and 2/3. (\textbf{b}) Repetition frequency map measured in the unlocked region between two Farey tree locking states with Farey fractions of 3/4 and 4/5. The modulation power from the RF generator is fixed at 30 dBm, and the tuning step size is 100 kHz. }
\label{periodicfrep}
\end{figure}

In addition to the locking bandwidth measurement, we also investigate the RF signal around the repetition frequency as a function of modulation frequency in unlocked regions and the results are shown in Fig. \ref{periodicfrep}. To accurately describe the influence of the microwave injection signal on the unlocked repetition frequency, we perform a measurement by finely tuning the modulation frequency and automatically monitoring the unlocked repetition frequency. In Fig. \ref{periodicfrep}a, between the two locking states with Farey fractions of 3/5 and 2/3, one can see a clear unlocked region from 3.6347 to 4.036 GHz. In the unlocked region, the repetition frequency shows a periodic oscillation with modulation frequency, and the period is measured to be 38.3 MHz. Similar oscillation and a period of 38.3 MHz are also observed for the unlocked region between 3/4 and 4/5 locking states, as shown in Fig. \ref{periodicfrep}b. It is worth noting that when we switch off the QCL and tune the modulation frequency, the measured power of the modulation frequency also demonstrates an oscillating behavior with the modulation frequency and the period is equal to 38.3 MHz again (see Supplementary Fig. S2a, Supporting Information). These oscillating behaviors with an identical period of 38.3 MHz observed in Fig. \ref{periodicfrep} and Supplementary Fig. S2a (Supporting Information) result from the non-ideal microwave transmission in the entire circuit loop due to the impedance mismatching. This causes the periodic change of the effective microwave power as the modulation frequency is varied under the same injection power level. Furthermore, the variations in injected power introduce strong current modulation and thermal effects, directly leading to a linear drift in the repetition frequency, as shown in Supplementary Fig. S2b (Supporting Information). Finally, the combined effects bring about periodic oscillations in the repetition frequency with modulation frequency. Because of this oscillating behavior, it is challenging to achieve the broadband Farey tree locking, and especially as the hierarchy deepens in the Farey tree the locking becomes more difficult.

\subsection{Farey tree locking plateaus}
\begin{figure}[tb]
\centering
\includegraphics[width=0.98\linewidth]{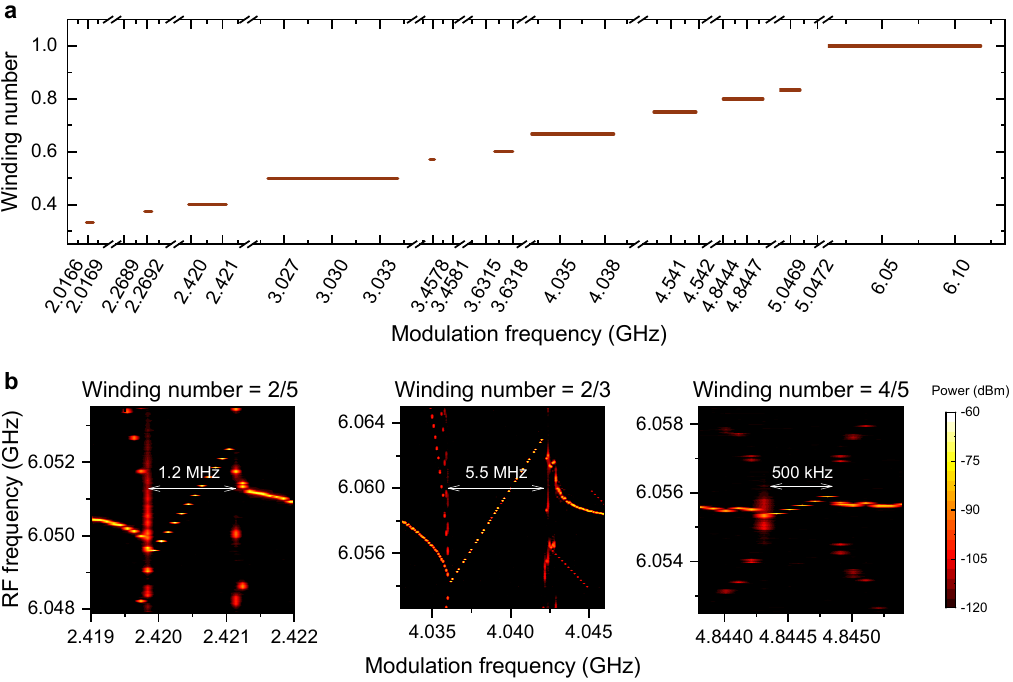}
\caption{Farey tree locking plateaus. (\textbf{a}) Winding number as a function of modulation frequency. The plateaus represent different Farey tree locking states when the winding numbers are equal to Farey fractions shown in Fig. \ref{farey tree}a. (\textbf{b}) Intermode beatnote maps under microwave modulation around different winding numbers of 2/5, 2/3, and 4/5. The tuning step size of modulation frequency is 100 kHz and the injection power is fixed at 30 dBm.}
\label{plateaus}
\end{figure}

The transition from unlocked to the Farey tree locking state is of great importance for understanding the dynamics of the locking process in THz QCL combs. In the following experiment, we again employ the automatic instrument control and data acquisition program. In this measurement, the modulation frequency starts from 2 GHz. As the modulation frequency is increased, the winding number increases accordingly. When the winding number is close to a Farey fraction, the Farey tree locking occurs. In the modulation frequency range between 2 GHz and $f_{\rm{rep}}$, 11 Farey tree locking plateaus are observed as shown in Fig. \ref{plateaus}a. The width of each plateau corresponds to the locking bandwidth at the given winding number (or Farey fraction). The measured locking bandwidths for the 11 plateaus are summarized in Supplementary Table S1 (Supporting Information). Similar to what we showed in Fig. \ref{Arnold}a, for the 11 plateaus, the locking bandwidth shrinks with the denominator of the winding number increases.

To clearly show the Farey tree locking dynamics, in Fig. \ref{plateaus}b the intermode beatnote maps under microwave modulation around different winding numbers of 2/5, 2/3, and 4/5 are shown. For each case, as the modulation frequency, $f_{\rm{mod}}$, increases, $f_{\rm{rep}}$ will be firstly pulled to lower frequencies; when $f_{\rm{mod}}$ is very close to $f_{\rm{rep}}$, side bands can be clearly observed, which refers to the strong interaction between the two frequencies, i.e., $f_{\rm{mod}}$ and $f_{\rm{rep}}$; then the Farey tree locking of $f_{\rm{rep}}$ is achieved, and in the locking bandwidth $f_{\rm{rep}}$ increases linearly with the increase of $f_{\rm{mod}}$ and the slope is equal to the inverse of the Farey fraction; finally, as $f_{\rm{mod}}$ is further increased beyond the locking range, $f_{\rm{rep}}$ can be no longer locked and broad peaks occur again. The left and right boundaries of the Farey tree locking range can be clearly observed as shown in Fig. \ref{plateaus}b and the measured locking bandwidths for Farey fractions of 2/5, 2/3, and 4/5 are 1.2 MHz, 5.5 MHz, and 500 kHz, respectively.

\subsection{Dual-comb operation under Farey tree locking condition}

\begin{figure}[!t]
\centering
\includegraphics[width=0.98\linewidth]{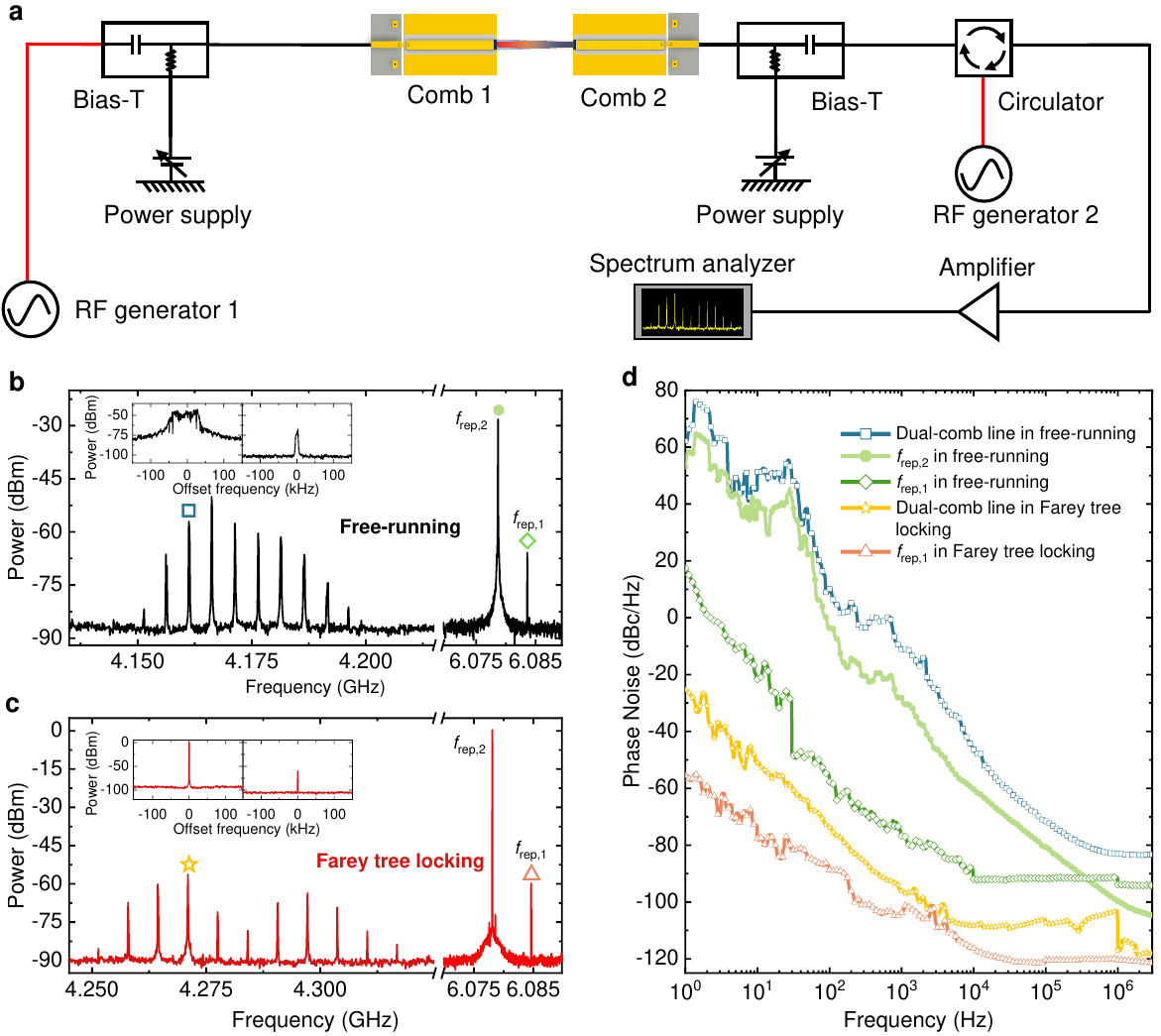}
%\captionsetup{font=small}
\caption{Dual-comb operation with Farey tree locking. (\textbf{a}) Experimental setup of the dual-comb measurement. Comb 1 and Comb 2 are mounted on a Y-shape sample holder. The Farey tree locking of Comb 1 and Comb 2 is achieved by modulating the two lasers at frequencies with Farey fractions of 2/3 (24 dBm) and 1/1 (-20 dBm), respectively. (\textbf{b}) and (\textbf{c}) Dual-comb and intermode beatnote spectra recorded in free-running and Farey tree locking states, measured using an RBW of 10 kHz and a VBW of 1 kHz. The insets show the corresponding high-resolution spectra of $f_{\rm{rep,2}}$ (left) and $f_{\rm{rep,1}}$ (right), measured using an RBW of 300 Hz and a VBW of 30 Hz. (\textbf{d}) Phase noise spectra of different lines marked in (b) and (c). The offset frequency range is set from 1 Hz to 3 MHz.}
\label{DCS}
\end{figure}

In order to further validate the effect of Farey tree locking, a dual-comb experiment is implemented. The schematic of the dual-comb experimental setup is shown in Fig. \ref{DCS}a. Two THz QCL frequency combs, i.e., Comb 1 and Comb 2, are mounted on a Y-shape cold finger\cite{li2020}. The two comb lasers are configured face to face and no mirrors are used for optical coupling between the two lasers. The two laser combs have identical nominal device dimensions, i.e., 150 $\mu$m wide ridge and 6 mm long cavity length. To operate the two lasers in frequency comb regimes, the applied drive currents are 990 and 930 mA for Comb 1 and Comb 2, respectively. The heat sink temperature is stabilized at 18.6 K. The multiheterodyne dual-comb signal resulting from the beating between Comb 1 and Comb 2 is measured using Comb 2 as a fast THz detector which shows a potential detection bandwidth upto 20 GHz\cite{guan2023}. The dual-comb spectra are finally displayed by a spectrum analyzer. The Farey tree locking with Farey fractions of 2/3 and 1/1 (resonant microwave injection locking) are applied to Comb 1 and Comb 2, respectively. In this experiment, the microwave power used for 2/3 Farey tree locking is 24 dBm, while it is set to be -20 dBm for 1/1 Farey tree locking to avoid large phase noise introduced by the resonant microwave injection\cite{liao2022}. 

Figure \ref{DCS}b shows the measured dual-comb spectrum and its corresponding intermode beatnotes ($f_{\rm{rep,1}}$ and $f_{\rm{rep,2}}$) when the two laser combs are operated in free-running mode. Since the entire signal is measured using Comb 2 as the detector, the intermode beatnote of Comb 1 ($f_{\rm{rep,1}}$) shows a much lower intensity than that of Comb 2 ($f_{\rm{rep,2}}$). The high-resolution spectra of free-running $f_{\rm{rep,1}}$ and $f_{\rm{rep,2}}$ are shown in the inset of Fig. \ref{DCS}b measured with an RBW of 300 Hz and a VBW of 30 Hz. In free-running mode, $f_{\rm{rep,1}}$ and $f_{\rm{rep,2}}$ show the frequency drift of 9 and 78 kHz, respectively. The free-running dual-comb spectrum clearly shows 10 lines with a line spacing of 5.1 MHz which is equal to the difference between the two repetition frequencies. Fig. \ref{DCS}c shows the dual-comb spectrum under the Farey tree locking condition. Under the Farey tree locking condition with Farey fractions of 2/3 and 1/1 for Comb 2 and Comb 1, respectively, the two intermode beatnotes $f_{\rm{rep,1}}$ and $f_{\rm{rep,2}}$ are firmly locked which can be proved by the high-resolution spectra shown in the inset of Fig. \ref{DCS}c. Benefiting from the Farey tree locking bandwidth, we are able to slightly tune the repetition frequency of the QCL combs, as shown in Supplementary Fig. S3 (Supporting Information). Compared to the dual-comb spectrum in free-running mode, we can see that under the Farey tree locking condition, the dual-comb bandwidth is not extended. This can be expected because the resonant microwave injection is relatively weak (-20 dBm) to avoid large phase noise.

Although the dual-comb bandwidth is not broadened by implementing the Farey tree locking, the stability of the dual-comb signal should be significantly improved. To systematically evaluate the stability of different signals, we perform the phase noise measurements by employing a phase noise analyzer (see Appendix). Figure \ref{DCS}d summarizes the phase noise spectra measured for different lines marked by scatters in Figs. \ref{DCS}b and \ref{DCS}c. First of all, it can be clearly seen that the locked signals, e.g., $f_{\rm{rep,1}}$ and dual-comb line in Farey tree locking, always demonstrate much lower phase noise than their counterparts measured in free-running mode. Specifically, at 10 Hz, 100 Hz, and 1 kHz, the measured phase noise levels of the Farey tree locked dual-comb line marked by the star in Fig. \ref{DCS}c are -50.9, -73.6, and -96.1 dBc/Hz, while the free-running dual-comb line marked by the square in Fig. \ref{DCS}b shows phase noise levels of 51.3, 10.1, and -11.2 dBc/Hz, respectively. More than 100 dBc/Hz improvement in phase noise is obtained at 10 Hz. The lower the offset frequency, the higher the improvement in the phase noise. To show the general effectiveness of the Farey tree locking, in Supplementary Fig. S4 more phase noise spectra of different dual-comb lines in free-running and Farey tree locking mode are shown. It is demonstrated that once the Farey tree locking is switched on, all dual-comb lines show improved phase noise levels than the dual-comb lines in free-running.

\section{DiSCUSSION}

In Figs. \ref{farey tree}-\ref{DCS}, we demonstrated that the Farey tree locking can be observed in THz QCL frequency combs and it can be further applied for the stabilization of dual-comb operation. Concerning the Farey tree locking method, there are several points to be emphasized. First of all, the Farey tree locking provides another alternative way to firmly lock the repetition frequency of THz QCL combs. Traditionally, the resonant microwave injection (a special Farey tree locking with a fraction of 1/1) has been widely used for repetition frequency stabilization of QCL frequency combs and the injected microwave frequency has to be equal to the fundamental repetition frequency of the comb laser. However, it is shown that the resonant microwave injection normally brings about high phase noise and then breaks frequency comb and dual-comb operation in THz QCLs\cite{li2019,liao2022}, especially when the injected microwave power is large. The Farey tree locking proposed in this work can overcome the noise problem. Because under the Farey tree locking condition the injection frequency is not at the repetition frequency of the comb laser, the noise at the Farey fraction frequency won't affect the comb and dual-comb operation. In view of this, large microwave injection power can be adopted to stabilize the repetition frequency and broaden the locking bandwidth. Secondly, compared to the resonant microwave injection at a fixed frequency, the Farey tree locking can be achieved at various Farey fraction frequencies, which provides many more choices for the frequency selection. Furthermore, by following the Farey tree structure shown in Fig. \ref{farey tree}a, the Farey fraction frequencies can be accurately anticipated. 

Apart from the advantages mentioned above, we have to point out that there is still room to improve the Farey tree locking technique. In principle, the fractal structure depicted in Fig. \ref{farey tree}a should extend to infinite deep layers. However, in this experiment, we can only reach a relatively shallow layer with a denominator of the Farey fraction of 8. One reason is that for the RF generator used in the measurements, the maximum output power is 30 dBm which strongly prevents us from reaching the deeper layers of the Farey tree structure. This can be clearly evidenced by the temporary locking observed for a Farey fraction of 5/7 (see Fig. \ref{farey tree}a). In this situation, due to the limited injection RF power, the laser is working in a temporary locking state, i.e., jumping between locked and unlocked states. Improvements can be obtained by adding a high power microwave amplifier to further increase the injection RF power onto the THz QCL chip. It can be expected that after the effective injection power is enhanced the Arnold tongue plots for 3/4 and 3/5 Farey fractions shown in Fig. \ref{Arnold}a will reveal the ``tongue" shape. On the other hand, it is worth noting that the impedance mismatching in the entire circuit loop leads to strong transmission line reflections (see Fig. \ref{periodicfrep}). The impedance mismatching results in the difficulty of achieving broadband Farey tree locking. Therefore, by increasing the microwave injection power and optimizing the impedance matching, extensions of the Farey tree structure can be revealed. Note that even under the current power and impedance mismatching conditions, we can explore the deeper fractal structure of the Farey tree by further reducing the frequency tuning step size. For example, as shown in Supplementary Fig. S5 (Supporting Information), when the step size is reduced to 20 Hz, the expected 4/11 plateau is observed. It is foreseeable that more fractal structures can be revealed with improved measurement accuracy.

Normally, the devil's staircase always accompanies the Farey tree. The Farey tree locking plateaus in Fig. \ref{plateaus} resemble the fractal structure of the devil's staircase, with step widths decreasing as the hierarchy of the Farey tree descends. By calculating the Cantor set formed by the plateaus in Fig. \ref{plateaus}a, the fractal dimension $D$ is determined to be 0.9985±0.0013 which shows a pronounced deviation from the theoretical fractal dimension of 0.87 for the devil's staircase\cite{HENTSCHEL1983,Jensen1983}. The deviation is originated from the fact that the measured plateau width shown in Fig. \ref{plateaus}a does not reach its maximum values due to the limited injected power and imperfect transmission of the RF signal due to the impedance mismatching.

In conclusion, we have demonstrated the Farey tree locking in THz QCL frequency combs under microwave injection. These Farey tree locking states were obtained by modulating the comb laser at Farey fraction frequencies which can be anticipated precisely. Under the Farey tree locking condition, the linewidth, signal-to-noise ratio, and phase noise of the repetition frequency were significantly improved, indicating a firm locking of the repetition frequency of the THz QCL comb. The observed locking plateaus corresponding to the Farey tree also reveal the existence of fractal structures in the THz QCL frequency comb. To further validate the Farey tree locking effectiveness, a dual-comb experiment was carried out. The experimental results showed that significant low phase noise of the dual-comb lines can be obtained by implementing the Farey tree locking technique: 100 dBc/Hz improvement in phase noise at an offset frequency of 10 Hz. The demonstrated Farey tree locking approach provides alternative ways to stabilize THz QCL frequency comb and dual-comb sources.

\begin{acknowledgments}

This work is supported by the Innovation Program for Quantum Science and Technology (2023ZD0301000), the National Science Fund for Distinguished Young Scholars (62325509), the National Natural Science Foundation of China (62235019, 61875220, 61927813, 61991430, 62035005, 62105351, and 62305364), the ``From 0 to 1" Innovation Program of the Chinese Academy of Sciences (ZDBS-LY-JSC009), and the CAS Project for Young Scientists in Basic Research (YSBR-069).

\end{acknowledgments}

\appendix
\section{THz QCLs}
THz QCLs employed in this work are based on a hybrid active region design that exploits the bound-to-continuum transition for THz photon emission and resonant phonon scattering for depopulation in the lower laser state for population inversion. The detailed layer structure of the proposed active region can be found in ref. [\citenum{wan2017}]. The entire active region of the QCL was grown on a semi-insulating GaAs (100) substrate using a molecular beam epitaxy (MBE) system. Then, the grown wafer was processed into single plasmon waveguide laser ridges with a ridge width of 150 $\mu$m. Finally, 6-mm-long laser bars were cleaved and indium-bonded onto copper heat sinks for wire bonding. The CW output power of the THz QCLs was measured with a THz power meter (Ophir, 3A-P-THz) using two parabolic mirrors for THz light collection and collimation. The power values shown in Fig. \ref{setup}b were displayed values on the screen of the power meter and no corrections were taken into account. During the power measurement, the lasers were operated in constant current mode.

\section{Farey tree}
The Farey tree is a specific sequence of rational numbers generated by applying the Farey-sum (or median) operation to two adjacent rational numbers. The Farey-sum operation is defined as follows:
\begin{equation}
\label{eq:median}
\frac{{m}}{{n}} \oplus \frac{{p}}{{q}} = \frac{{m+p}}{{n+q}}
\end{equation}
Here, the variables $m$, $n$, $p$, and $q$ represent positive natural numbers. A lower-hierarchy Farey fraction is generated from the two parental nodes of the upper-hierarchy. The downward generation process results in the self-similarity between local and global levels of the Farey tree. Based on the derivation of the Farey-sum, an infinite number of Farey fractions exist between any two Farey fractions. In theory, an infinite number of Farey fractions correspond to an infinite number of Farey tree locking plateaus, which form the devil's staircase of a fractal structure. The locking bandwidth gradually decreases within the devil's staircase due to its fractal nature.

\section{Farey tree locking technique}
As shown in Fig. \ref{setup}a, the Farey tree locking was carried out by modulating the THz QCL comb at a Farey fraction frequency. RF generator (Anritsu, MG3693C) with a frequency range from 8 MHz to 31.8 GHz and a power level from -20 to 30 dBm was utilized to provide a strong RF signal to modulate the QCL's current for locking. The locking state of the repetition frequency of the THz QCL was monitored using a spectrum analyzer (Rohde \& Schwarz, FSW26).

For the Farey tree locking of the THz QCL dual-comb source, as shown in Fig. \ref{DCS}a two locks with Farey fractions of 2/3 and 1/1 were implemented. For the 2/3 Farey tree locking, RF generator 1 identical to the one used in Fig. \ref{setup}a was employed to lock Comb 1; for 1/1 Farey tree locking, RF generator 2 (Rohde \& Schwarz, SMA100B) with a frequency range from 8 kHz to 20 GHz and a power level from -90 to 20 dBm was used to lock Comb 2. A microwave circulator (Ditom, D3C5964) operating in the frequency range between 5.9 and 6.4 GHz with a minimum isolation of 23 dB was used for the directional injection and detection of RF signals. The intermode beatnotes, as well as the multiheterodyne dual-comb signals, were measured using the device itself as a detector and finally registered on a spectrum analyzer (Rohde \& Schwarz, FSW26). The phase noise spectra shown in Fig. \ref{PN}c, Fig. \ref{DCS}d, and supplementary Fig. S4c were measured using a phase noise analyzer (Rohde \& Schwarz, FSWP26). For the phase noise measurement of low power ($<$-40 dBm) intermode beatnote and dual-comb signals, the function of ``low-pass filter" with a cutoff frequency of 10 MHz was switched on to reduce the influence of the broadband noise.

\bibliography{apsref}% Produces the bibliography via BibTeX.

\end{document}